\documentclass[aps,pre,showpacs,showkeys,onecolumn]{revtex4-1}
\usepackage{graphicx,amsmath,amsfonts,amssymb,dsfont}
\usepackage[activeacute,english]{babel}
\newcommand{\weiave}[1]{\left\langle \hspace{-2.5pt}\left\langle #1 \right\rangle \hspace{-2.5pt}\right\rangle}
\newcommand{\sgn}{{\rm sgn}}

\begin{document}

\title{Stochastic functionals and fluctuation theorem for the multi-kangaroo process}

\author{C. Van den Broeck}
\affiliation{ Hasselt University, B-3590 Diepenbeek, Belgium}
\author{R. Toral}

\affiliation{ IFISC (Instituto de F{\'\i}sica Interdisciplinar y Sistemas Complejos), Campus UIB, Palma de Mallorca, Spain}

\begin{abstract}{ We introduce multi-kangaroo Markov processes and provide a general procedure for evaluating a certain type of stochastic functionals. We calculate analytically the large deviation properties. Applications include zero-crossing statistics and stochastic thermodynamics. }
\end{abstract}

\keywords{Kangaroo process, stochastic functionals, stochastic thermodynamics, fluctuation theorem}

\pacs{05.70.Ln;05.20.-y;05.40.-a}
\maketitle
\section{Introduction}

The theory of stochastic processes provides a powerful general setting to describe the dynamical behavior of systems, including their fluctuations. In many cases, one is interested in stochastic functionals that depend on the entire history of the process. It is then often difficult, or even impossible, to obtain a full analytical solution. One therefore focuses on simple examples, such as the Ising model which, with its many equilibrium and nonequilibrium variations and extensions, is one of the best studied and most productive paradigms in statistical mechanics \cite{isingmodel}. Its relative simplicity derives from the fact that the basic ingredient is a two-state spin system. For a single spin, the dynamics thus corresponds to an alternation between these two states and the corresponding transition probabilities are defined in terms of the internal interactions between the spin variables as well as with external thermal baths. 
An alternative simplifying assumption that reduces drastically the mathematical complexity of the problem is to assume that whenever the system changes its state, it samples its available states according to an a-priori prescribed probability distribution, independent of the current state. Such a dynamics has been termed a {\sl kangaroo process} \cite{vankampen}.  One significant advantage with respect to the Ising-type of models is that it can accomodate a general spectrum of states. 
In the context of Markovian stochastic process, a kangaroo is characterized by a transition rate that factorizes in terms of the variables of the initial and final states. This model has been extensively applied in various contexts, including kinetic theory \cite{kt}, line shape analysis \cite{briso}, mathematical biology \cite{mb} and turbulence \cite{turb}.
Various generalizations of the kangaroo process have been considered in the literature, including non-Markovian variants \cite{nonmarkov}. 

In this paper, we introduce a generalization of the kangaroo process, which we refer to as ``multi-kangaroo'', with the transition rate being the sum of factorizable contributions. This new model carries a much richer physical content, while keeping, as we will see, the mathematical simplicity of the single-component kangaroo process. One reason for distinguishing different constitutive kangaroo rates is that each may correspond to a different physical process, which we want to identify separately, or which have a different effect in the environment. Another motivation is the recently developed theory of stochastic thermodynamics. In this theory, the stochastic thermodynamic properties of nonequilibrium states can be investigated. The nonequilibrium state can be realized by simultaneous exposure to different equilibrium environments, which -as we will see- correspond to the separate dynamics of the multi-kangaroo process. The quantities of interest in this theory are stochastic functionals such as the entropy production or heat flow, which are usually very difficult to evaluate \cite{maes}. We will show that their calculation is enormously simplified for multi-kangaroo processes. In particular their large deviation properties are obtained as the largest eigenvalue of a finite matrix. In this context, we stress that the standard literature on large deviations in Markov processes is focusing on functionals like the empirical distribution \cite{ldf}, whereas we are dealing with quantities that depend, not only on the trajectory in state space, but also on the type of process that is responsible for changes in the state of the system. 

The paper is organized as follows: after introducing the multi-kangaroo process in section \ref{multi-kangaroo}, we show in section \ref{stochastic-functionals} how the probability distribution of stochastic functionals can be obtained by solving a finite set of coupled linear differential equations. In section \ref{large-deviations} we focus on the large deviation properties of these functionals and show that the asymptotic cumulant generating function is the largest eigenvalue of the finite matrix appearing in the aforementioned linear system. As applications we first compute in section \ref{net-zero-crossings} the asymptotic cumulant generating function of the cumulated net zero-crossings associated to one of the processes. Second we evaluate, in section \ref{fluctuation-theorem},  the asymptotic stochastic entropy production and verify that the famous fluctuation theorem is obeyed \cite{stochtherm}. Explicit expressions for the statistical properties of the entropy production are given in section \ref{model-systems} for an  energy spectrum  corresponding to  single particle and quasi-particle states and for the quantum harmonic oscillator. We end in section \ref{perspectives} with some conclusions and perspectives for further studies.

\section{Multi-kangaroo process}
\label{multi-kangaroo}
A Markovian kangaroo process is a Markov process in which the transition rate to go from state $x'$ to $x$, $W(x'\rightarrow x)$, factorizes in terms of the variables of the initial and final states: $W(x'\rightarrow x)=\phi(x')\rho(x)$ \cite{vankampen}. We will restrict ourselves here to the case of $\phi(x')$ being independent of $x'$, because this condition is required for a consistent description in stochastic thermodynamics, cf. section \ref{fluctuation-theorem}. The transition rate thus has the following form: 
\begin{equation}\label{1}
W(x'\rightarrow x)=k P_\textrm{ss}(x).
\end{equation}
The physical interpretation is as follows: $k$ is the constant rate at which transitions take place; whenever a transition occurs, the new state $x$ is chosen from the probability density $P _\textrm{ss}(x)$, which is in fact the steady state distribution of the Markov process. To fix the ideas, we will assume in the following that $x$ is a real variable, but other and in particular more abstract interpretations are possible. 
 The master equation for the probability $P(x;t)$ to be in state $x$ at time $t$:
\begin{equation}\label{2}
\partial_t P(x;t)=\int\; dx\;' [W(x'\rightarrow x)P(x';t)-W(x\rightarrow x')P(x;t)]
\end{equation}
simplifies as follows for the kangaroo process defined by (\ref{1}):
\begin{equation}\label{3}
\partial_t P(x;t)=-k[P(x;t)-P _\textrm{ss}(x)].
\end{equation}
We conclude that the ``kangaroo master equation'' (\ref{2}) describes a pure exponential relaxation (with a characteristic decay time $k^{-1}$), towards the steady state $P _\textrm{ss}(x)$. 

We now generalize the kangaroo process as follows: a transition from $x'$ to $x$ can be realized by several distinct physical processes $\nu$, each with corresponding rates $W^{(\nu)}(x'\rightarrow x)$. The resulting total rate is given by: 
\begin{equation}
W(x'\rightarrow x)=\sum_{\nu}W^{(\nu)}(x'\rightarrow x).
\end{equation}
 To keep the simplicity of the kangaroo process, we assume that each of the rates has the same form as in (\ref{1}):
 \begin{equation}\label{7}
W^{(\nu)}(x'\rightarrow x)=k^{(\nu)} P^{(\nu)} _\textrm{ss}(x).
\end{equation}
Here, $k^{(\nu)}$ is the (constant) rate of transitions associated to process $\nu$, and $P^{(\nu)} _\textrm{ss}$ the corresponding steady state distribution. The master equation corresponding to the resulting total rate $W(x'\to x)$ is still given by (\ref{3}), but with the following total rate $k$ and overall steady state distribution $P _\textrm{ss}$:
\begin{equation}\label{8}
k=\sum_\nu k^{(\nu)},\hspace{20pt}
P _\textrm{ss}(x)=\frac{\sum_\nu k^{(\nu)}P^{(\nu)} _\textrm{ss}(x)}{\sum_\nu k^{(\nu)}}.
\end{equation}
Hence, at this level of description solely it terms of the state $x$, there is no difference with the usual kangaroo process, if at least one decides not to discriminate between the processes that are responsible for the transitions. In the following we will however consider stochastic functionals  that make this distinction. The above subtle difference between the coarse-grained and detailed process is a familiar situation in stochastic processes \cite{langevin}. For example an overdamped Brownian particle in contact with different heat baths can be described by an effective Langevin equation in which the different contributions are lumped together. If the distinction between the baths is not made, a nonequilibrium state is mistaken for an equilibrium one.

\section{Stochastic functionals}
\label{stochastic-functionals}
We want to evaluate the probability distribution for an ``incremental cumulative'' quantity $\Delta=\Delta(t)$ associated to the stochastic trajectory generated by the Markovian stochastic process over a time interval of length $t$. A simple example is the cumulative number of transitions: the value of $\Delta$ changes by $+1$ whenever a transition takes place, i.e., the increment of $\Delta$ is $\delta(x'\rightarrow x)=1$, independent of $x$ and $x'$. Another example is the net number of ``net zero-crossings" (or flux through $0$), corresponding to $\delta(x'\rightarrow x)=[\sgn(x)-\sgn(x')]/2$, with $\sgn(x)$ representing the sign of $x$. As a third example we mention the cumulative energy exchanged between a system and a bath, with $\delta(x'\rightarrow x)=\epsilon(x)-\epsilon(x')$, and $\epsilon(x)$ the energy of the system in state $x$.

For the multi-kangaroo process, the increments of $\Delta$ will depend on the states between which the transition takes place, but also on the process responsible for them, i.e., the increments are given by $\delta^{(\nu)}(x'\rightarrow x)$ for a transition $x'\rightarrow x$ due to process $\nu$. One example is the entropy production for a system in contact with different heat baths at temperature $T^{\nu}$, namely $\delta^{(\nu)}(x'\rightarrow x)=[\epsilon(x)-\epsilon(x')]/T^{(\nu)}$. Another example is the cumulated effect due to one of the processes, say $\nu_0$, implying $\delta^{(\nu)}(x'\rightarrow x)$ is zero for all processes except for $\nu=\nu_0$.
 
In all the above cited examples, the increments have a specific feature in common: they can be written as the sum (or difference) of a function of $x$ and $x'$, i.e.:
\begin{equation}\label{incr}
 \delta^{(\nu)}(x'\rightarrow x)=b^{(\nu)}(x)-a^{(\nu)}(x').
 \end{equation}
It turns out that this feature, combined with the kangaroo property of the transition rate, greatly reduces the mathematical complexity of the problem, as we now proceed to show.

Since the increases of $\Delta$ are supposed to be a deterministic function of the transitions, the combined pair $x\,,\Delta$ obeys a master equation, which obviously reads as follows:
\begin{eqnarray}\label{ld1}
\partial_t P(x,\Delta;t)=\sum_{\nu}\int\; dx\;' [W^{(\nu)}(x'\rightarrow x)P(x',\Delta-\delta^{(\nu)}(x'\rightarrow x);t)-W^{(\nu)}(x\rightarrow x')P(x,\Delta;t)].\nonumber\\
\end{eqnarray}
It is convenient to introduce the following generating function of the $\Delta$ variable:
 \begin{equation}\label{ld2}
P_\lambda(x;t)=\int\; d\Delta \;e^{\lambda \Delta} P(x,\Delta;t),
\end{equation}
which obeys the following evolution equation:
\begin{equation}\label{ld3}
\partial_t P_\lambda(x;t)=\sum_{\nu}\int\; dx'\; [W^{(\nu)}(x'\rightarrow x)e^{\lambda \delta^{(\nu)}(x'\rightarrow x)} P_\lambda(x';t)-W^{(\nu)}(x\rightarrow x')P_\lambda(x;t)].
\end{equation}
It is now clear why a significant simplification takes place for the multi-kangaroo scenario considered here. By taking the transition rates and increments given by (\ref{7}) and (\ref{incr}), respectively, the above equation (\ref{ld3}) reduces to:
\begin{equation}\label{ld4}
\partial_t P_\lambda(x;t)=\sum_\nu k^{(\nu)} e^{\lambda b^{(\nu)} (x)}P^{(\nu)} _\textrm{ss}(x) I^{(\nu)}_\lambda(t) -kP_\lambda(x;t).
\end{equation}
Here, we introduced the integrals $I^{(\nu)}_\lambda(t)$:
\begin{equation}
I^{(\nu)}_\lambda(t)=\int\; dx\; e^{-\lambda a^{(\nu)}(x)} P_\lambda(x;t).\end{equation}
The quantity of prime interest is the cumulant generating function:
\begin{equation}
F_\lambda(t)=\langle e^{\lambda \Delta}\rangle=\exp\left({\sum_{k=0}^\infty \frac{\lambda^k \weiave{\Delta^k}}{k!}}\right)=\int dx \int d\Delta\; e^{\lambda \Delta} P(x,\Delta;t)=\int\; dx\; P_\lambda(x;t),
\end{equation}
where $\weiave{\Delta^k}$ is the cumulant of order $k$.
Combination of the former three equations leads to the following closed set of linear equations for $F_\lambda(t)$ and the $I^{(\nu)}_\lambda(t)$:
\begin{eqnarray}\label{ldm1}
\partial_t F_\lambda(t) &=& \sum_{\nu'} k^{(\nu')} B^{(\nu')}_\lambda I^{(\nu')}_\lambda(t)- k F_\lambda(t),\\
\partial_t I^{(\nu)}_\lambda(t) &=&\sum_{\nu'} k^{(\nu')}A^{(\nu,\nu')}_\lambda I^{(\nu')}_\lambda(t)-k I^{(\nu)}_\lambda(t).\label{ldm12}
\end{eqnarray}
Here $A^{(\nu,\nu')}_\lambda$ and $B^{(\nu)}_\lambda$ are the following time-independent quantities:
\begin{eqnarray}\label{A}
A^{(\nu,\nu')}_\lambda=\int\; dx\; e^{\lambda[b^{(\nu')}(x)-a^{(\nu)}(x)]} P^{(\nu')} _\textrm{ss}(x),\hspace{10pt}
 B^{(\nu)}_\lambda=\int\; dx\; e^{\lambda b^{(\nu)}(x)} P^{(\nu)} _\textrm{ss}(x).
\end{eqnarray}
Note that the above set of equations (\ref{ldm1}) and (\ref{ldm12}) can be written under a matrix form:
\begin{eqnarray}
{\bf{\dot V}}_\lambda(t)={\bf{M}}_\lambda{\bf{V}}_\lambda(t),
\end{eqnarray}
 where the vector ${{\bf{V}}_\lambda}(t)$ has components $F_\lambda(t),I^{(1)}_\lambda(t),...,I^{(N)}_\lambda(t)$, $N$ being the number of processes $\nu$. ${\bf{M}}_\lambda$ is a time-independent $(N+1)\times(N+1)$ matrix, whose elements can be identified by inspection of the equations (\ref{ldm1}) and (\ref{ldm12}). Furthermore the $I_\lambda$ components do not couple to $F_\lambda$, so that their dynamics is ruled by a time-independent $N\times N$-matrix with coefficients  $k^{(\nu')}A^{(\nu,\nu')}_\lambda-k\;\delta^{(\nu,\nu')}_{\rm Kr}$. The exact, but formal, analytic solution ${\bf{V}}_\lambda(t)={\rm e}^{t {\bf{M}}_\lambda}{\bf{V}}_\lambda(0)$ can be made explicit for particular values of the matrix coefficients  or, quite generally, for  any functional involving a small number, say $N=2$ or $N=3$, of processes. In conclusion, the evaluation of stochastic functionals is reduced to the discussion of the dynamics induced by a finite matrix. The size of this matrix is equal to the number of processes, and in particular independent of the spectral density, i.e. of the type or number of possible states $x$. 

\section{Large deviations}
\label{large-deviations}
We next focus on the large deviation properties of $\Delta(t)$ in the asymptotic limit $t \rightarrow \infty$. As the Markov process $x$ does not possess long-time correlations, the increments of $\Delta$ cumulated over time periods longer than the correlation time, are essentially independent. Hence the behaviour of $\Delta(t)$ for ${t\rightarrow \infty}$ is typically described by the asymptotic cumulant generating function $\phi_\lambda$ \cite{ldf}:
\begin{equation}\label{ldc}
\phi_\lambda=\lim_{t\rightarrow \infty}\frac{\ln\langle \exp\left({\lambda \Delta(t)}\right)\rangle}{t},\;\;\;{\rm or}\;\;\;F_\lambda(t)=\langle \exp\left({\lambda \Delta(t)}\right)\rangle\sim_{t\rightarrow \infty} \exp\left(t\phi_\lambda\right).
\end{equation}
 Alternatively, one can focus on the asymptotic form of the probability distribution $P(\Delta;t)$. For $t \rightarrow \infty$ the current $j(t)=\Delta(t)/t$ will converge by the law of large numbers to its average value, which also corresponds to its most probable value (assuming that the latter is unique). The exponentially rare deviations of the current from this average value are described by the large deviation function $\psi_j$ : 
\begin{equation}
P(\Delta=jt;t)\sim_{t\rightarrow \infty} \exp\left(-t\psi_j\right).
\end{equation}
Application of Laplace's theorem to the cumulant generating function leads to the conclusion that 
$\psi_j$ is the Legendre transform of the asymptotic cumulant generating function (Varadhan's theorem, \cite{ldf}):
\begin{equation}\label{ltr}
\psi_j=\textrm{ext}_{\lambda}\left(\lambda j-\phi_\lambda\right),
\end{equation}
If we assume that $\phi_\lambda$ is differentiable in $\lambda$, the transform can be obtained by inverting $j=\phi'_\lambda$ to obtain $\lambda=\lambda(j)$ and then replace in $\psi_j=j\lambda(j)-\phi_{\lambda(j)}$.

Comparing (\ref{ldc}) to the formal solution ${\bf{V}}_\lambda(t)={\rm e}^{t {\bf{M}}_\lambda}{\bf{V}}_\lambda(0)$, it is clear that $\phi_\lambda$ has to be identified with the largest eigenvalue of the matrix ${\bf {M}}_\lambda$. 
The analysis is, as already mentioned, further simplified by the observation that the equations for the $I_\lambda^{(\nu)}$ components do not couple to $F_\lambda$. In particular, we identify the eigenvalue equal to $-k$ associated to the $F_\lambda$ component, i.e, corresponding to the eigenvector $(1,0,...,0)^\intercal$. $\phi_\lambda$ is the largest eigenvalue of the block-matrix related to the $I_\lambda^{(\nu)}$ components whose elements we identified as $k^{(\nu')}A^{(\nu,\nu')}_\lambda-k\;\delta^{(\nu,\nu')}_{\rm Kr}$. In conclusion $\phi_\lambda+k$ is the largest eigenvalue of the $N\times N$ matrix with elements $k^{(\nu')}A^{(\nu,\nu')}_\lambda$.

For the case of a single process, $N=1$, we can drop the superscripts $\nu$ and $\nu'$ in the above formulas. 
 We conclude that $\phi_\lambda=k(A_\lambda -1)$ (note that $A_\lambda\geq 0$), or explicitly:
\begin{eqnarray}\label{ldm}
\phi_\lambda=k \int\; dx\; \left(e^{\lambda[a(x)-b(x)]}-1\right) P _\textrm{ss}(x).
\end{eqnarray}
This result can also be obtained directly by noting that the number $n$ of transitions during a time $t$ obeys a Poisson distribution and that the contributions $\delta^{(i)}=b(x^{(i)})-a(x^{(i)})$ ($x^{(i)}, i=0,...,n$ being the successive states of the system) are independent and identically distributed random variables with probability distribution given by $P(\delta)=\int dx \delta_{\textrm{Dirac}}(\delta-b(x)+a(x))P _\textrm{ss}(x) $. 

For the case of two processes, the asymptotic cumulant generating function is found to be: 
\begin{eqnarray}\label{ldm2}
\phi_\lambda=\frac{-2k+k^{(1)}A_\lambda^{(1,1)}+k^{(2)}A_\lambda^{(2,2)}+\sqrt{\left(k^{(1)}A_\lambda^{(1,1)}-k^{(2)}A_\lambda^{(2,2)}\right)^2+4k^{(1)} k^{(2)} A_\lambda^{(1,2)}A_\lambda^{(2,1)}}}{2}.\nonumber\\
\end{eqnarray}
The asymptotic cumulant generating function can also be explicitly obtained for the general case with three processes using Cardano's formula for the roots of a third-degree polynomial, but the expression is too lengthy to be reproduced here. 

\section{Net zero-crossings}
\label{net-zero-crossings}
For the special choice $a^{(\nu)}(x)=b^{(\nu)}(x)\equiv q(x)$, $\forall \nu$, one finds $A^{(\nu,\nu')}_\lambda\equiv 1$, implying $\phi_\lambda=0$, and all normalized cumulants $t^{-1}\weiave{\Delta^k}$ vanish in the long time limit. This is however no longer the case when $a^{(\nu)}(x)=b^{(\nu)}(x)=q^{(\nu)}(x)$, $\forall \nu$, but with functions $q^{(\nu)}(x)$ that are not identical. As an illustration, we evaluate the asymptotic cumulant generating function for the cumulated net zero-crossings of a process $\nu=1$, in the presence of another ``resetting" process $\nu=2$, whose zero-crossings are not counted. More precisely we set:
\begin{equation}
a^{(1)}(x)=b^{(1)}(x)=\sgn(x)/2,\;\;\;\;a^{(2)}(x)=b^{(2)}(x)=0.
\end{equation}
The asymptotic cumulant generating function is given by (\ref{A}), with the following values:
\begin{eqnarray}
&&A_\lambda^{(1,1)}=A_\lambda^{(2,2)}=1,\;\;\;
A_\lambda^{(1,2)}=e^{-\lambda} P^{(2)}_++e^{\lambda}P^{(2)}_-,\;\;\;
A_\lambda^{(2,1)}=e^{\lambda}P^{(1)}_++e^{-\lambda}P^{(1)}_-,\nonumber\\
&&P^{(\nu)}_{\pm}=\int_{\sgn(x)=\pm} dx\;\;\; P _\textrm{ss}^{(\nu)}(x).
\end{eqnarray}
The result becomes particularly transparent for equal rates $k^{(1)}=k^{(2)}=k/2$ and $ P^{(1)}_\pm=P^{(2)}_\pm=1/2$, namely
$\phi_\lambda=k(e^{\lambda}+e^{-\lambda}-2)/4$, which is the asymptotic cumulant generating function of an unbiased random walk with jump rate $k/4$. Indeed, the probability to be in $+$ or $-$ state is equal to $1/2$ at all times, and the probability per unit time to select process $(1)$ for a jump is $k^{(1)}=k/2$, hence $k/4$ is the rate of zero-crossings by process $(1)$ for both $+ \rightarrow -$ and $- \rightarrow +$.

\section{Fluctuation theorem}
\label{fluctuation-theorem}
As second example, we consider a system in contact with different heat baths $\nu=1,\dots,N$
with corresponding temperatures $T^{(\nu)}$. The transitions between different states are due to the contact with the heat baths. In particular, a transition $x'\rightarrow x$ requires the following amount of heat $Q(x'\rightarrow x)=\epsilon(x)-\epsilon(x')$, being $\epsilon(x)$ the energy of the system when in state $x$.
If this transition is produced by contact with heat bath $\nu$, the corresponding entropy change in the bath is given by (minus sign as we are monitoring the entropy change of the bath):
\begin{equation}
\delta^{(\nu)}(x'\rightarrow x)= -\frac{Q(x'\rightarrow x)}{T^{(\nu)}}=\frac{\epsilon(x')-\epsilon(x)}{T^{(\nu)}}.
\end{equation}
This corresponds to the choice:
\begin{equation}
\label{anubnu}
a^{(\nu)}(x)=b^{(\nu)}(x)= -\frac{\epsilon(x)}{T^{(\nu)}}.
\end{equation}
The sum $\Delta$ of all the contributions $\delta^{(\nu)}$ is, for any realization of the process, equal to the total stochastic entropy production in the reservoirs.

With respect to the application of stochastic thermodynamics \cite{stochtherm}, we note that the kangaroo transition rates, associated to bath $\nu$, automatically satisfy the required condition of (local) detailed balance, $W^{(\nu)}(x \rightarrow x') P^{(\nu)} _\textrm{ss}(x)=W^{(\nu)}(x' \rightarrow x) P^{(\nu)} _\textrm{ss}(x')$, because we assumed that the rates $k^{(\nu)}$ do not depend on the state. The corresponding steady distribution $P^{(\nu)} _\textrm{ss}$ should however also reproduce the equilibrium distribution when in contact with this bath, hence it is given by:
\begin{equation}\label{canonical}
P^{(\nu)} _\textrm{ss}(x)=P^{(\nu)}_\textrm{eq}(x)=\frac{\exp[-\beta^{(\nu)}\epsilon(x)]}{Z(\beta^{(\nu)})},
\end{equation}
where we introduced the partition function $Z(\beta)$:
\begin{equation}
\label{zcanonical}
Z(\beta)=\int dx \exp[-\beta\epsilon(x)]= \int d\epsilon g(\epsilon) \exp(-\beta\epsilon).
\end{equation}
Here $g(\epsilon)=dx/d\epsilon$ is the density of states and $\beta$ has the usual definition $1/\beta=k_B T$. 
With the identification (\ref{anubnu}), one finds that (\ref{A}) simplifies as follows:
\begin{eqnarray}\label{A1}
A^{(\nu,\nu')}_\lambda&=&\int\; dx\; e^{-\lambda \epsilon(x)[1/T^{(\nu')}-1/T^{(\nu)}]} P^{(\nu')}_{eq}(x)=\frac{Z\left(k_B\lambda[\beta^{(\nu')}-\beta^{(\nu)}]+\beta^{(\nu')}\right)}{Z(\beta^{(\nu')})}.
\end{eqnarray}

We now proceed to the proof of the so-called fluctuation theorem.  We mentioned before that $\phi_\lambda+k$ is the largest eigenvalue of the matrix of coefficients $\tilde A^{(\nu,\nu')}=k^{\nu'}A^{(\nu,\nu')}_\lambda$. For the characteristic polynomial of this $N\times N$ matrix we use the expansion in minors:
\begin{equation}
{\textrm{Det}}[x\mathds{1}-\tilde A]=\sum_{i=0}^{N}(-1)^i\sigma_{i}x^{N-i}
\end{equation}
with $\sigma_i$ being the sum over all the $i$-th order diagonal minors of matrix $\tilde A$. Now we note from (\ref{A1}) that the elements of the matrix $A$ satisfy the following symmetry property: 
\begin{equation}
{Z(\beta^{(\nu')})}{A^{(\nu,\nu')}_\lambda}={Z(\beta^{(\nu)})}{A^{(\nu',\nu)}_{-\lambda-1/k_B}}.
\end{equation}
This implies that in the calculations of the minors contributing to $\sigma_i$ we can use
\begin{equation}
\prod\limits_{\nu\in S} {\tilde A^{(\nu,\cal{P}(\nu))}_\lambda}=\prod\limits_{\nu\in S} {\tilde A^{(\nu,\cal{P}(\nu))}_{-\lambda-1/k_B}},
\end{equation}
where $S$ is any subset of $(1,2,...,N)$ and $\cal{P}$ any permutation of this subset. This implies that the characteristic polynomial (and hence its roots) is invariant under the transformation $\lambda\rightarrow-\lambda-1/k_B$. The cumulant generating function $\phi_\lambda$, related to the largest of the eigenvalues, inherits this property:
\begin{equation}
\label{fluctphi}
\phi_\lambda=\phi_{-\lambda-1/k_B}.
\end{equation} 
This implies by Legendre transformation, cf. (\ref{ltr}), the following symmetry behavior for $\psi_j$, and the corresponding asymptotic behavior for $\Delta=jt$:
\begin{equation}
\psi_j-\psi_{-j}=-\frac{j}{k_B}\;\;\;\mbox{and}\;\;\;\frac{P(\Delta)}{P(-\Delta)}\sim \exp\frac{\Delta}{k_B}.
\end{equation}
Since $\Delta$ is the cumulated entropy production in the reservoirs, the above result is nothing but the celebrated asymptotic or steady state fluctuation theorem \cite{stochtherm,early}. 

By expanding $A^{(\nu,\nu')}_\lambda$ is powers of $\lambda$,  cf. (\ref{A1}), one can show that  $\phi_\lambda=-k+\lambda\phi_1+O(\lambda^2)$ with  
\begin{equation}
\label{averagen}
\phi_1=\lim_{t\rightarrow \infty}\frac{\langle {\Delta}\rangle}{t}=\sum_{\nu'<\nu}\left(\frac{1}{T^{(\nu)}}-\frac{1}{T^{(\nu')}}\right)\frac{k^{(\nu)}k^{(\nu')}}{k}\left(\langle\epsilon\rangle^{(\nu')}-\langle\epsilon\rangle^{(\nu)}\right),
\end{equation}
where $\langle\cdots\rangle^{(\nu)}$ is the average calculated with respect to the canonical distribution (\ref{canonical}). This formula expresses the total entropy production as the sum of the corresponding contributions for all possible channels $(\nu\leftrightarrow \nu')$.

In the case of two reservoirs $\nu=1,2$, besides the explicit results for the asymptotic average rate of entropy production that we quote here for later reference:
\begin{equation}
\label{average}
\lim_{t\rightarrow \infty}\frac{\langle {\Delta}\rangle}{t}=\left(\frac{1}{T^{(2)}}-\frac{1}{T^{(1)}}\right)\frac{k^{(1)}k^{(2)}}{k^{(1)}+k^{(2)}}\left(\langle\epsilon\rangle^{(1)}-\langle\epsilon\rangle^{(2)}\right),
\end{equation}
one can also derive the fluctuations
\begin{equation}
\label{sigma}
\lim_{t\rightarrow \infty}\frac{\sigma^2[\Delta]}{t}=2\left.\frac{\partial^2\phi_\lambda}{\partial \lambda^2}\right|_{\lambda=0}=\left(\frac{1}{T^{(2)}}-\frac{1}{T^{(1)}}\right)^2\frac{k^{(1)}k^{(2)}}{k^{(1)}+k^{(2)}}\left(\sigma^2[\epsilon]^{(1)}+\sigma^2[\epsilon]^{(2)}+\frac{{k^{(1)}}^2+{k^{(2)}}^2}{(k^{(1)}+k^{(2)})^2}\left(\langle\epsilon\rangle^{(1)}-\langle\epsilon\rangle^{(2)}\right)^2\right),
\end{equation}
where $\sigma^2[\epsilon]^{(\nu)}=\langle\epsilon^2\rangle^{(\nu)}-(\langle\epsilon\rangle^{(\nu)})^2$.
For the case of small temperature difference $T^{(2)}-T^{(1)}$ one can expand $\langle\epsilon\rangle^{(2)}=\langle\epsilon\rangle^{(1)}+C(T^{(1)})(T^{(2)}-T^{(1)})+O((T^{(2)}-T^{(1)})^2$, $C(T)$ being the specific heat, to obtain
\begin{equation}
\label{average2}
\lim_{t\rightarrow \infty}\frac{\langle {\Delta}\rangle}{t}=\frac{k^{(1)}k^{(2)}}{k^{(1)}+k^{(2)}}C(T^{(1)})\eta_C^2+O(\eta_C^3),
\end{equation}
$\eta_C=1-T^{(1)}/T^{(2)}$ being the Carnot efficiency. Using Einstein's relation $\sigma^2=k_BT^2C(T)$ we derive at the same order
\begin{equation}
\label{sigma2}
\lim_{t\rightarrow \infty}\frac{\sigma^2[\Delta]}{t}=2k_B\frac{k^{(1)}k^{(2)}}{k^{(1)}+k^{(2)}}C(T^{(1)})\eta_C^2+O(\eta_C^3).
\end{equation}
The expansion of both cumulants to $O(\eta_C^3)$ thus reproduces the Gaussian linear response regime.

\section{Model systems}
\label{model-systems}
\begin{figure}
\begin{center}
\includegraphics[width=8cm]{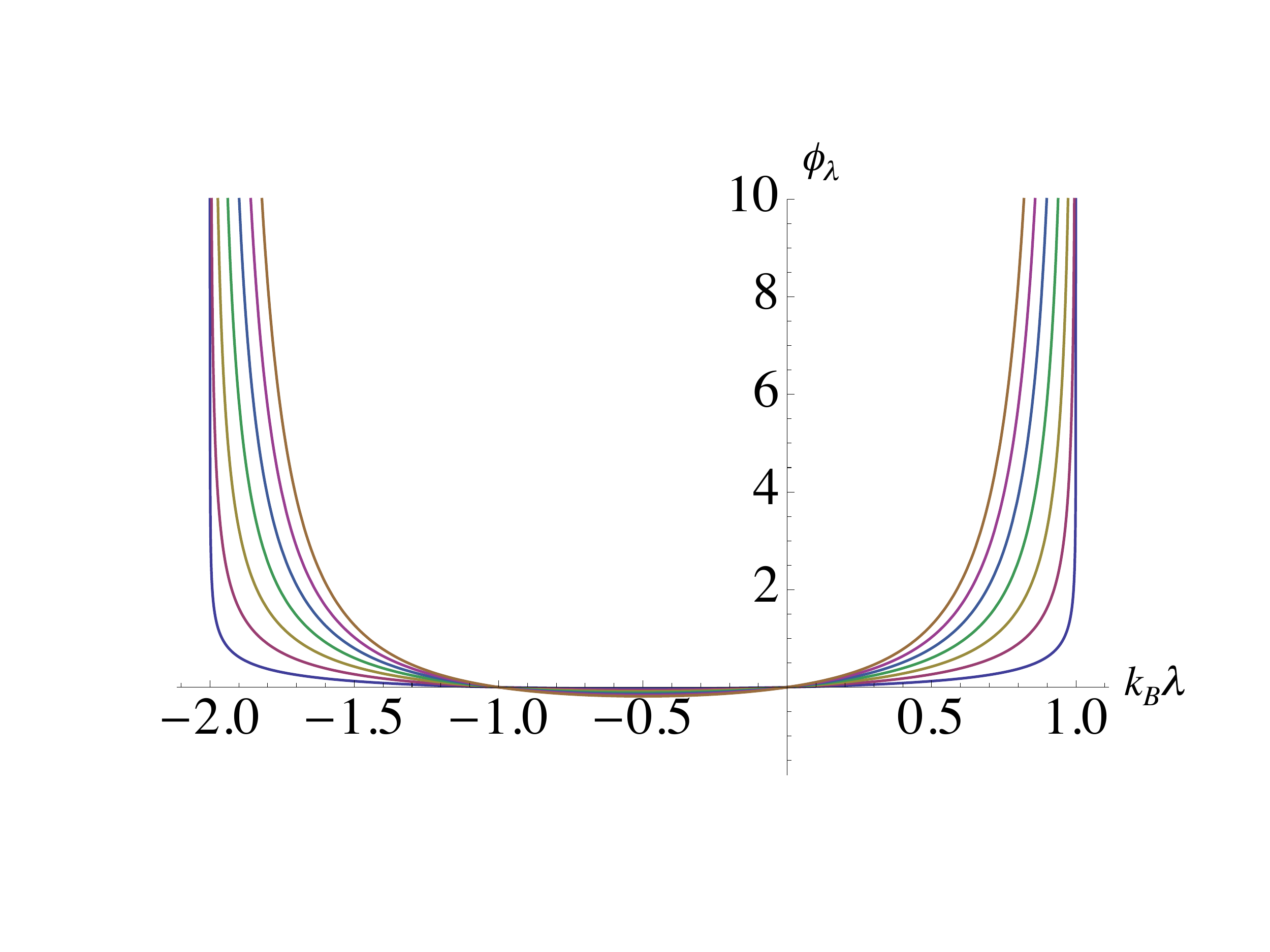}\includegraphics[width=8cm]{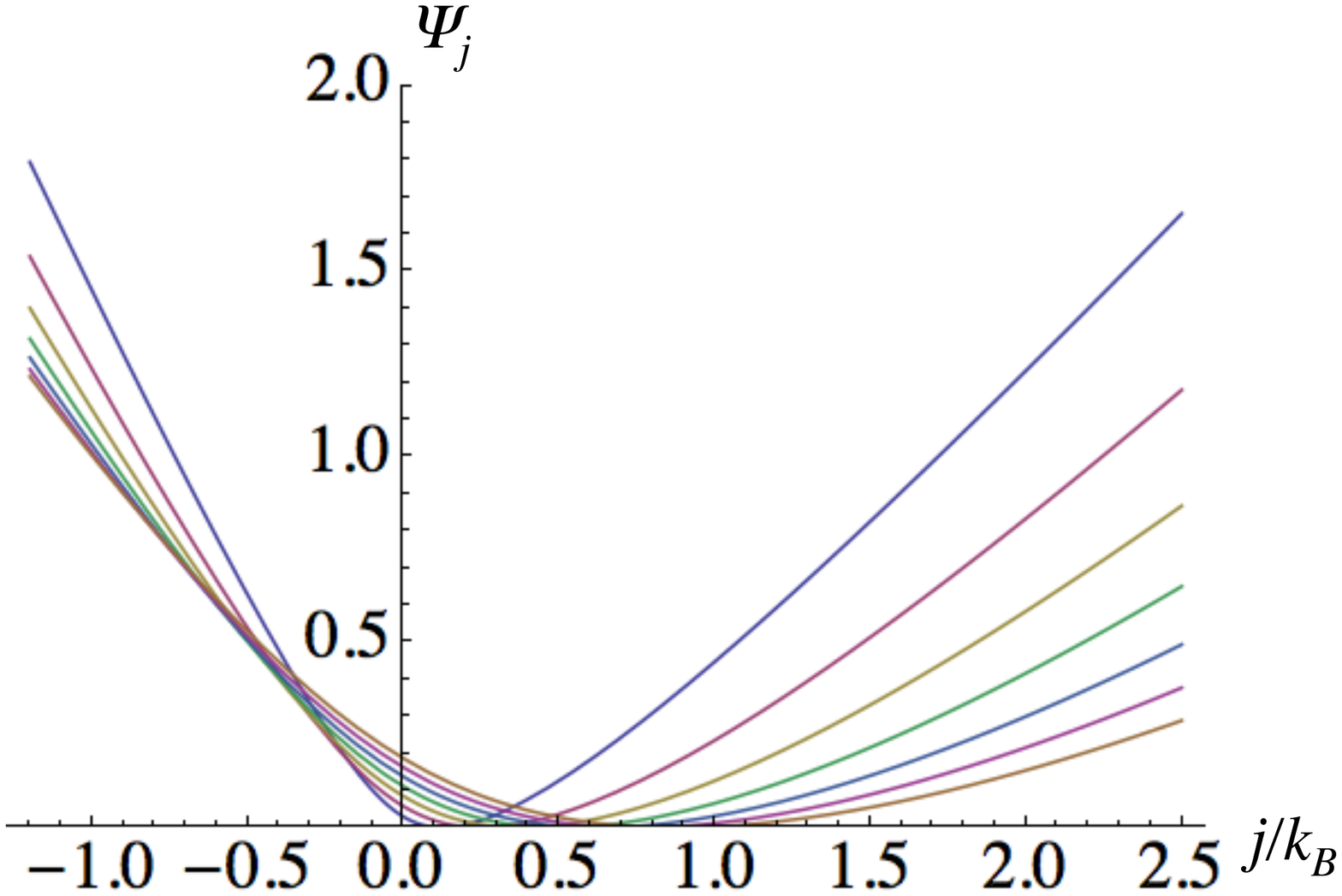}
\end{center}
\caption{(color online) Asymptotic cumulant generating function $\phi_\lambda$ (left panel) and the corresponding large deviation function $\psi_j$ (right panel) for the energy density $g(\epsilon)\sim (\epsilon-\epsilon_0)^{\alpha-1}$, $\epsilon>\epsilon_0$. $\alpha$ goes from $\alpha=1/2$ (outer curve in left panel; upper curve for positive abscissas in the right panel) to $\alpha= 7/2$ in steps of $1/2$, with $\eta_C=1/2$. The symmetry of $\phi_\lambda$ about the point $k_B\lambda=-1/2$, imposed by the fluctuation theorem, is clearly visible . Note also the divergences at $ k_B \lambda=-1/\eta_C=-2$ and $k_B \lambda=1/\eta_C-1=1$.}
\label{fig1}
\end{figure}

To proceed further, one needs to specify the density of states $g(\epsilon)$ of the system under consideration. The simplest case corresponds to a discrete spectrum with two energy states $\epsilon=0$ and $\epsilon=\epsilon_0$. The $\lambda$ dependence of $\phi_\lambda$ is then similar to the one in a general two-state problem, which is discussed in detail in \cite{np}. Since the strength of the Kangaroo model is that it can be applied to a general spectrum, we focus here on more complicated energy spectra that are relevant in solid state physics. A general dispersion relation of the form $\epsilon=\epsilon_0+ap^b$ covers most cases of interest for particles or quasi-particles \cite{Kittel}. The density of states of the momentum in $d$ spatial dimensions is given by $g(p)=C_dp^{d-1}$, with $C_d=\displaystyle 2\pi^{d/2}V/(\Gamma(d/2)h^d)$,  $V$ the volume and $h$ Planck's constant. We thus obtain  $g(\epsilon)=g_0 (\epsilon-\epsilon_0)^{\alpha-1}$, $\epsilon>\epsilon_0$, with $g_0=\displaystyle C_d/(ba^{\alpha})$ and $\alpha=d/b$. In this case, the calculations can be done analytically. The partition function (\ref{zcanonical}) reads: 
\begin{equation}
Z(\beta)=g_0\Gamma(\alpha)e^{-\beta \epsilon_0}\beta^{-\alpha},
\end{equation}
and the asymptotic cumulant generating function is given by, cf. (\ref{ldm2}) and (\ref{A1}):
\begin{equation}
\phi_{\lambda}=-\frac{k}{2}+\sqrt{\left(\frac{k^{(1)}-k^{(2)}}{2}\right)^2+k^{(1)}k^{(2)}\left[(1-\frac{\eta_C}{1-\eta_C}k_B\lambda)(1+\eta_Ck_B\lambda)\right]^{-\alpha}}.
\end{equation}
Note that $\phi$ depends only on the ratio of temperatures. This can be understood from the fact that there is no other energy in the model, $\epsilon_0$ being solely a lower bound of the energy spectrum. One furthermore verifies that $\phi_\lambda$ is symmetric around $k_B\lambda=-1/2$, as imposed by the fluctuation theorem (\ref{fluctphi}). The cumulant generating function $\phi_\lambda$ is plotted in Fig. \ref{fig1} for a few representative situations, together with its Legendre transform $\psi_j$. Note the divergences of   $\phi_\lambda$ at the borders of the interval in which it is defined, $k_B\lambda\in]-1/\eta_C,1/\eta_C-1[$  (assuming $T^{(2)}>T^{(1)}$). This is a general feature of systems with an unbounded energy spectrum, as the effective inverse temperature parameter in the partition function, $k_B\lambda[\beta^{(\nu')}-\beta^{(\nu)}]+\beta^{(\nu')}$, has to be positive for $\nu'=1, \nu=2$ and $\nu'=2,\nu=1$.  

For completeness, we mention also the results for the first two cumulants. The average and variance of the energy are equal to: 
\begin{eqnarray}
\langle \epsilon\rangle^{(\nu)} &=&\epsilon_0+\alpha k_BT^{(\nu)} \\
\sigma^2[\epsilon]^{(\nu)}&=&\alpha (k_BT^{(\nu)})^2.
\end{eqnarray}
From (\ref{average}-\ref{sigma}) we obtain that the first two cumulants of the entropy production rate are given by:
\begin{eqnarray}
\lim_{t\rightarrow \infty}\frac{\langle {\Delta}\rangle}{t}&=&\alpha k_B \frac{k^{(1)}k^{(2)}}{k^{(1)}+k^{(2)}}\frac{\eta_C^2}{1-\eta_C}\\
&=&\alpha k_B \frac{k^{(1)}k^{(2)}}{k^{(1)}+k^{(2)}}\eta_C^2+O(\eta_C^3),
\end{eqnarray}
and 
\begin{eqnarray}
\lim_{t\rightarrow \infty}\frac{\sigma^2[\Delta]}{t}&=&\alpha k_B^2\frac{k^{(1)}k^{(2)}}{k^{(1)}+k^{(2)}}\frac{\eta_C^2}{(1-\eta_C)^2}\left(2(1-\eta_C)+\left(1+\alpha\frac{{k^{(1)}}^2+{k^{(2)}}^2}{(k^{(1)}+k^{(2)})^2}\right)\eta_C^2\right)\\
&=&2\alpha k_B^2\frac{k^{(1)}k^{(2)}}{k^{(1)}+k^{(2)}}\eta_C^2+O(\eta_C^3).
\end{eqnarray}

We next present the results for the harmonic oscillator with energy levels $\epsilon_n=(n+\frac12)\hbar\omega$ with its well-known  partition function:
\begin{equation}
Z(\beta)=\frac{2}{\sinh\left(\frac12\beta\hbar\omega\right)}.
\end{equation}
 The asymptotic cumulant generating function is found to be, cf. (\ref{ldm2}) and (\ref{A1}):
\begin{equation}
\phi_\lambda=-\frac{k}{2}+\sqrt{\left(\frac{k^{(1)}-k^{(2)}}{2}\right)^2+2k^{(1)}k^{(2)}\frac{\sinh\left(\frac12\hbar\omega\beta^{(1)}\right)\sinh\left(\frac12\hbar\omega\beta^{(2)}\right)}{\cosh\left(\frac12\hbar\omega(\beta^{(1)}+\beta^{(2)})\right)-\cosh\left(\frac12\hbar\omega(2k_B\lambda+1)(\beta^{(1)}-\beta^{(2)})\right)}}.
\end{equation}
It verifies the required symmetry $\phi_\lambda=\phi_{-\lambda-1/k_B}$ and  is defined in the interval $k_B\lambda\in]-1/\eta_C,1/\eta_C-1[$, diverging to $+\infty$ at both ends. The average energy is:
\begin{equation}
\langle \epsilon \rangle^{(\nu)}=\frac12\hbar\omega\left[1+\frac{e^{-\frac12\hbar\omega\beta}}{\sinh\left(\frac12\hbar\omega\beta\right)}\right].
\end{equation}
The average rate of entropy production reads: 
\begin{eqnarray}
\lim_{t\rightarrow \infty}\frac{\langle {\Delta}\rangle}{t}&=& k_B \frac{k^{(1)}k^{(2)}}{k^{(1)}+k^{(2)}}\hbar\omega(\beta^{(2)}-\beta^{(1)})\frac{\sinh\left(\frac12\hbar\omega(\beta^{(2)}-\beta^{(1)})\right)}{2\sinh\left(\frac12\hbar\omega\beta^{(2)}\right)\sinh\left(\frac12\hbar\omega\beta^{(1)}\right)}\\
&=&k_B \frac{k^{(1)}k^{(2)}}{k^{(1)}+k^{(2)}}\left[\frac{\frac12\hbar\omega\beta_1}{\sinh\left(\frac12\hbar\omega\beta_1\right)}\right]^2\eta_C^2+O(\eta_C^3)
\end{eqnarray}

\section{Perspectives}
\label{perspectives}
We have presented some illustrations of the large deviation theory for a generalized ``kangaroo scenario".
It should however be clear that the model can be applied to a wide range of problems. The variables $x$ could be vectors (for example the speed of a particle or several variables such as energy and number of particles), functions or fields (for example probability distributions, density or flux profiles), matrices or operators (with the quantity of interest for example its largest eigenvalue), or more abstract quantities (for example symbols or processes), with the corresponding probability distributions $P _\textrm{ss}$, processes $\nu$ and increments $\delta$ having widely different interpretations. 
While the true stochastic dynamics will typically be more involved, the analysis of a generalized kangaroo model may lead to analytic results that can serve as a guideline for the properties of the original system. It also provides an alternative to a description in terms of a two-state Ising-type model, with which it shares the mathematical simplicity, while keeping the richness associated to an arbitrary spectrum of states and steady state distribution.

\section{Acknowledgments}
This work was supported by the research network program ``Exploring the Physics of Small Devices" from the European Science Foundation. RT acknowledges financial support from MINECO (Spain), and FEDER (EC) under project FIS2012-30634 and Comunitat Aut\`onoma de les Illes Balears,

\end{document}